\documentclass[12pt]{article}		
\usepackage[super,sort,comma]{natbib}

\usepackage{fancyhdr}		

\newcommand{\captionv}[3]{\begin{center}\parbox{#1cm}{\caption[#2]{{\sf #3}}}
        \end{center}}

\usepackage[section]{placeins}   %

\usepackage{graphicx}
\graphicspath{ {./figures/} }

\makeatletter \renewcommand\@biblabel[1]{$^{#1}$} \makeatother
\renewcommand{\thesubsection}{\thesection{\sf \Alph{subsection}}.}

\newcommand{\cen}[1]{\begin{center} #1 \end{center}}

\usepackage{hyperref}
\hypersetup{ colorlinks,
    citecolor=blue,
    filecolor=blue,
    linkcolor=blue,
    urlcolor=blue
}

\usepackage{xcolor}

\definecolor{gray}{rgb}{0.6,0.6,0.6}
\definecolor{red}{rgb}{0.85,0,0}
\definecolor{green}{rgb}{0,0.85,0}
\definecolor{blue}{rgb}{0,0,0.85}
\definecolor{beige}{rgb}{0.92,0.87,0.78}
\usepackage[all]{hypcap}    

\usepackage{todonotes}
\usepackage{amsmath,amssymb,mathtools}
\usepackage{algorithm}
\usepackage[noend]{algpseudocode}
\usepackage{bm}
\usepackage{makecell}
\usepackage{changepage}

\begin{document}

\begin{titlepage}
\thispagestyle{empty}

\cen{\sf {\Large {\bfseries A parallel algorithm for generating\\Pareto-optimal radiosurgery treatment plans}} \\  
\vspace*{1mm}
{\large Joakim da Silva\footnote{Author to whom correspondence should be addressed. Email: joakim.dasilva@elekta.com}$^1$, Daniel Hernández Escobar$^2$,\\Tor Kjellsson Lindblom$^1$, Håkan Nordström$^1$, Jens Sjölund$^2$} \\[1mm]
{ $^1$ Elekta Instrument AB, Stockholm, Sweden\\
$^2$ Department of Information Technology, Uppsala University, Sweden}
}

\begin{adjustwidth}{-3mm}{-3mm}
\begin{abstract}
\noindent{\bf Background:} Using inverse planning tools to create radiotherapy treatment plans is an iterative process, where clinical trade-offs are explored by changing the relative importance given to different objectives and rerunning the optimizer until a desirable plan is found. Simultaneously generating many plans corresponding to different objective weights, while the patient is awaiting treatment, would allow the planner to navigate clinical trade-offs interactively, without optimizing a new plan between each update. \\ 
{\bf Purpose:} We seek to optimize hundreds of radiosurgery treatment plans, corresponding to different weightings of objectives, fast enough to allow interactive Pareto navigation of clinical trade-offs to be incorporated into the clinical workflow.\looseness=-1 \\
{\bf Methods:} We apply the alternating direction method of multipliers (ADMM) to the linear-program formulation of the optimization problem used in Lightning. We implement both a CPU and a GPU version of ADMM in Matlab and compare them to Matlab’s built-in, single-threaded dual-simplex solver. The ADMM implementation is adapted to the optimization procedure used in the clinical software, with a bespoke algorithm for maximizing the overlap between low-dose points for different objective weights. The method is evaluated on a test dataset consisting of 20 cases from three different indications, with between one and nine targets and total target volumes ranging from 0.66 to 52 cm\textsuperscript{3}.\\
{\bf Results:} The total optimization time to create 81 plans corresponding to different objective weightings varied from 63 to 520 seconds on CPU and from 1.8 to 40 seconds GPU, for the different test cases. As a reference, optimizing 81 plans using simplex took 100--51000 seconds, corresponding to ADMM speedups of 1.6--97 and 54--1500 times for the CPU and GPU, respectively. Increasing the number of plans to 441, corresponding to all combinations of slider values between 0.0 and 1.0 in steps of 0.05 in the clinical software, the total ADMM optimization time on GPU was between 3.0 and 110 seconds for the different test cases.

Plan quality was evaluated by rerunning the ADMM optimization 20 times, each with a different random seed, for each test case and for nine objective weightings per case. The resulting relative differences in clinical metrics (mean±SD) were 0.0±0.2\%, 0.0±1.6\%, 0.1±0.8\%, and 0.1±3.0\%, for coverage, selectivity, gradient index and beam-on time, respectively, compared to mean values for the corresponding reference simplex results. The standard deviations in these metrics closely mimicked those obtained when rerunning the simplex solver, verifying the validity of the method. \\
{\bf Conclusions:} We show how ADMM can be adapted for radiosurgery plan optimization, allowing hundreds of high-quality Gamma Knife treatment plans to be created in under two minutes on a single GPU, also for very large cases. The presented method would allow streamlined multicriteria optimization on the day of treatment, with interruption-free navigation of clinical trade-offs.
\end{abstract}
\end{adjustwidth}





\end{titlepage}

\pagenumbering{arabic}
\setcounter{page}{1}
\pagestyle{fancy}
\section{Introduction}
Radiosurgery using the Leksell Gamma Knife (Elekta AB, Stockholm, Sweden) is a non-invasive treatment modality designed for high-precision irradiation of intracranial lesions. It delivers focused radiation through multiple cobalt-60 sources, with all beams converging at a precisely defined isocenter. Sources are grouped together in so-called sectors, where the collimation of the beams from different sectors is controlled independently. Treatment plans are constructed from a sequence of discrete ``shots'', each characterized by an isocenter location, a collimator selection for each sector, and an irradiation time. Combining multiple shots enables the generation of conformal dose distributions tailored to complex target geometries.\looseness=-1

Recent advances in inverse planning for the Gamma Knife, notably the Lightning module of Leksell GammaPlan (Elekta AB, Stockholm, Sweden), have demonstrated that high-quality treatment plans can be generated automatically in under a minute \citep{sjolund2019linear}. This capability is enabled by an efficient formulation as a large-scale linear program, representing a shift from manual, trial-and-error-based planning toward highly automated optimization-driven workflows\cite{hamavckova2025comparative}. The clinical goals of radiosurgery are often inherently conflicting, such as maximizing tumor coverage while minimizing dose to surrounding healthy structures and ensuring treatment efficiency\cite{craft2006approximating}. In the Lightning software, the planner explores trade-offs between clinical goals by adjusting two sliders, controlling the dose to healthy tissue (``low-dose'' slider) and treatment time (``beam-on-time'' slider), and rerunning the optimization.

The conflicting goals naturally lead to the formulation as a multicriteria optimization problem. In linear multicriteria optimization, weighted-sum formulations can be used to recover the Pareto front, defined as the set of all non-dominated solutions where improvement in one criterion necessitates degradation in another\cite{romeijn2004unifying}. Conventional multicriteria optimization workflows often assume either \textit{a priori} preference articulation\cite{breedveld2012icycle,tolakanahalli2023optimal}, where objective weights are specified before optimization, or \textit{a posteriori} navigation\cite{miettinen2008introduction,monz2008pareto}, where the clinician explores trade-offs after solving a set of representative plans. While the former approach offers speed, it is often difficult to encode clinical preferences with sufficient fidelity. The latter, in contrast, provides more intuitive control but can be computationally burdensome, as each new plan requires a separate optimization\cite{bokrantz2013algorithm}.
This is especially challenging in the radiosurgery workflow, where planning must often be completed on the day of treatment, between imaging and treatment delivery, placing stringent demands on computational efficiency.

To support rapid, high-throughput planning workflows, we explore the use of the Alternating Direction Method of Multipliers (ADMM)\citep{Glowinski,Gabay}. ADMM splits the optimization variables into subsets that are optimized alternately, and often lends itself well to parallel implementation. Specifically, we adapt an ADMM-based optimization solver designed for parallel GPU execution and tailored to the structure of the radiosurgical planning problem. This includes a novel method for efficiently handling objectives related to reducing dose spillage, which would otherwise prevent optimizing for different preferences in parallel. Our approach enables simultaneous generation of hundreds of Pareto-optimal treatment plans within minutes, demonstrating that multicriteria planning with explicit low-dose control can be achieved at clinically relevant runtimes.

\section{Methods}

\subsection{Problem Formulation}\label{sec:formulation}
We follow the inverse planning framework of Sjölund et al.\cite{sjolund2019linear}, where the aim is to promote plans with high coverage and high selectivity, while at the same time promoting a low gradient index and a short beam-on time. Here, coverage is defined as the fraction of the target volume that receives at least the prescription dose, selectivity as the fraction of the volume receiving at least the prescription dose that intersects the target, and gradient index as the volume receiving at least half the prescription dose divided by the volume receiving at least the prescription dose. The beam-on time is simply the total irradiation time required to deliver the plan.

By treating the isocenter positions as fixed, we can recast the problem as a multicriteria linear program with four objectives penalizing (i) dose below the prescription inside the target (to promote high coverage), (ii) dose above the prescription outside the target (to promote high selectivity), (iii) dose above a selected low‐dose threshold outside the target (to promote low gradient index), and (iv) total beam‐on time (to promote short treatment time). We introduce the weight vector
$
(w_\text{T},\,w_\text{R},\,w_\text{LD},\,w_{\text{BOT}})\in\mathbb{R}_{+}^4
$
to control their relative importance via a weighted sum. Hard constraints on the maximum dose to any point in an organ at risk (OAR) complete the formulation.

Although there are many potential methods of isocenter selection\cite{wu2003real,doudareva2015skeletonization,ghobadi2012automated}, we rely here on the set of isocenters produced by the clinical Lightning software, where the isocenter placement is a function only of target geometry. This approach has been shown to provide sufficient flexibility to achieve good plan metrics and allow clinically relevant trade-offs based on the assigned objective weights, whilst keeping optimization times sufficiently short\cite{hamavckova2025comparative}. Importantly, the selection of objective weights does not affect the selection of isocenters.

Similarly to Sjölund et al.\cite{sjolund2019linear}, for computational efficiency, we sample our dose points from volumes of interest, rather than using all voxel doses in the optimization. Specifically, dose points for the different objectives are selected from: (i) the surface and interior of the target volume, (ii) a ring just outside the target surface, and (iii) one or more low-dose regions further away from the target. Where OARs are present, dose points for the corresponding max-dose constraints are sampled from their surfaces, which is sufficient under the assumption that no isocenter position is placed inside an OAR. While more sophisticated sampling strategies exist \cite{fountain2022dose,mair2024efficient,quarz2024deep}, and may reduce the size of the optimization problem further, we do not expect these to impact the conclusions of our work. 

To exploit the problem structure, it is computationally favorable to solve the dual problem\citep{sjolund2019linear}, which is of the form
\begin{equation}  \label{eq:dual_problem}
\begin{aligned}
& \underset{x}{\text{minimize}}
& & c^\top x \\
& \text{subject to}
& & Ax \leq b,\\
& & & \ell \leq x \leq u.
\end{aligned}
\end{equation}
For readability, we defer the radiosurgery-specific definitions and parameter choices to Appendix~\ref{app:dual_lp}, but remark that most of the physics goes into the matrix $A\in\mathbb{R}^{m\times n}$ while the weights of the primal problem described above enter as components in the upper bounds, $b$ and $u$, in the dual formulation.

\subsection{A parallel solver based on ADMM}\label{sec:admm}
Our work is motivated by the practical need to solve the planning problem for a large ensemble of weight vectors to approximate the Pareto surface and thereby enable real-time clinical decision-making. (Recall that each weight vector corresponds to a unique trade-off between clinical criteria.)
Naively solving hundreds of LPs sequentially is inefficient and highly time-consuming, as we show in Section ~\ref{sec:timing}.
Instead, what is needed is a method that can handle batch processing efficiently. 

We implement an optimization algorithm based on the Alternating Direction Method of Multipliers (ADMM)\citep{boyd2011distributed}. ADMM splits the optimization variables into two sets of primal variables that are accompanied by dual variables. In brief, the solver iteratively updates the primal and dual variables to minimize the augmented Lagrangian, with each update step parallelized across weight vectors. To set the stage,
let 
\begin{equation}
\mathcal{I}_\mathcal{X}(x)=\begin{cases}
        0 & x\in\mathcal{X},\\
        +\infty & \text{otherwise}
    \end{cases}    
\end{equation}
be the \emph{indicator function} of the set $\mathcal{X}$ and
$\mathcal{A} = \{
    (x_1, x_2) \in \mathbb{R}^n \times \mathbb{R}^m \mid Ax_1 = x_2\}$.
We rewrite problem \eqref{eq:dual_problem} by introducing ``copies" $z$ of the optimization variable $x$ and moving inequalities into the objective function as follows
\begin{equation}  \label{eq:dual_problem_admm}
\begin{aligned}
& \underset{x_1,x_2,z_1,z_2}{\text{minimize}}
& & c^\top x_1 +\mathcal{I}_{\mathcal{A}}(x_1, x_2) + \mathcal{I}_{[\ell, \,u]}(z_1) + \mathcal{I}_{(-\infty, \,b]}(z_2)\\
& \text{subject to}
& & x_1 = z_1,\\
& & & x_2 = z_2.
\end{aligned}
\end{equation}
Though seemingly artificial, the split above, which is due to Nair et al.\cite{nair2020solving}, allows us to use ADMM to solve our problem. Specifically, this split yields the following three steps for the ADMM algorithm when expressed in scaled form\cite{boyd2011distributed}:
\begin{subequations}\label{eq:admm}
\begin{align}
\begin{pmatrix}
x_1^{k+1} \\[1mm]
x_2^{k+1}
\end{pmatrix}
&= \underset{(x_1, x_2) \in \mathcal{A}}{\arg\min}
\left\{
c^\top x_1 + \frac{\rho}{2} \left\| 
\begin{pmatrix}
x_1 - z_1^k + y_1^k \\[1mm]
x_2 - z_2^k + y_2^k
\end{pmatrix}
\right\|_2^2
\right\}, \label{eq:x-step}\\[1em]
\begin{pmatrix}
z_1^{k+1} \\[1mm]
z_2^{k+1}
\end{pmatrix}
&=
\begin{pmatrix}
\Pi_{[\ell, \,u]}\left(x_1^{k+1} + y_1^k\right) \\[1mm]
\Pi_{(-\infty, \,b]}\left(x_2^{k+1} + y_2^k\right)
\end{pmatrix}, \label{eq:z-step} \\[1em]
\begin{pmatrix}
y_1^{k+1} \\[1mm]
y_2^{k+1}
\end{pmatrix}
&=
\begin{pmatrix}
y_1^k + x_1^{k+1} - z_1^{k+1} \\[1mm]
y_2^k + x_2^{k+1} - z_2^{k+1}
\end{pmatrix},
\end{align}
\end{subequations}
where the superscripts $k$ denote the iteration number and $y$ are the scaled dual variables, which are associated with an augmented Lagrangian parameter $\rho$ controlling the trade-off between primal and dual feasibility (and coinciding with the step size in the dual variable update\cite{boyd2011distributed}). In equation \eqref{eq:z-step}, the Euclidean projection
$\Pi_\Omega(x)=\underset{v\in\Omega}{\arg\min \,} \|x-v\|_ 2$ simplifies to element-wise clipping.
By iterating the above ADMM steps, the primal variable $z_1^k$ converges to the solution $x^*$ of the  problem in equation \eqref{eq:dual_problem}. Simultaneously, the scaled dual variable $y_2^k$
approaches a vector $y^*$. The solution of our original (primal) problem, including the sector times,
is $\rho y^*$.

The computational bottleneck in the ADMM algorithm of equation \eqref{eq:admm} is solving the equality-constrained least-squares problem in the first step. Fortunately, this can be reduced to solving the following system of equations\cite{nair2020solving}:
\begin{equation}\label{eq:eq_system}
\begin{pmatrix}
I & A^\top \\[1mm]
A & I
\end{pmatrix}
\begin{pmatrix}
x_1^{k+1} \\[1mm]
x_2^{k+1} - z_2^k + y_2^k
\end{pmatrix}
=
\begin{pmatrix}
z_1^k - y_1^k - c/\rho \\[1mm]
z_2^k - y_2^k
\end{pmatrix}.
\end{equation}
Solving for the second component, $v^k \coloneq x_2^{k+1} - z_2^k + y_2^k$, yields
\begin{equation}
v^k = (AA^\top + I)^{-1} \left( A (z_1^k - y_1^k - c/\rho) - z_2^k + y_2^k \right),    \label{eq:partial_elimination}
\end{equation}
where $S = AA^\top + I$ is the Schur complement of the top-left submatrix of the left-hand side in equation \eqref{eq:eq_system}. 
Importantly, $S$ does not change from one iteration to another. We can, therefore, compute its inverse (or decomposition)  just once and cache it. In subsequent iterations, equation \eqref{eq:partial_elimination} then amounts to a matrix-vector multiplication. In particular, we can replace equation \eqref{eq:x-step} with the closed-form expression 
\begin{equation}    
\begin{pmatrix}
x_1^{k+1} \\[1mm]
x_2^{k+1}
\end{pmatrix}
=
\begin{pmatrix}
z_1^k - y_1^k - c/\rho - A^\top v^k \\[1mm]
v^k + z_2^k - y_2^k
\end{pmatrix}.
\end{equation}
Hence, the ADMM algorithm in \eqref{eq:admm} consists solely of matrix-vector multiplications, vector additions, and component-wise projections of vectors. Not only does this make each iteration computationally cheap---it also allows parallelization over the different weight vectors in the primal problem, which enter only through the upper bounds $b$ and $u$ of equation~\eqref{eq:z-step}. The only change required to solve $n$ instances in parallel is to stack the instance-specific vectors (i.e., $x_k^i$, $y_k^i$, $z_k^i$, $l$, $u$, and $b$) as columns, and use the resulting matrices instead of the vectors in \eqref{eq:admm}.

For a given problem, this allows us to find a large number of Pareto-optimal points by running the ADMM algorithm once, where each iteration consists only of matrix-matrix multiplications and additions, and element-wise projections---operations that lend themselves to very efficient implementation on parallel compute architectures such as GPUs.

Two additional computational aspects related to $S$ are worth noting. First, $S^{-1}$ does not have to be computed explicitly. Instead, it might be advantageous to compute and subsequently cache a suitable decomposition of $S$, which can then be used to repeatedly solve equations of the form $S v^k = b$. If we work with, for instance, an LU or Cholesky decomposition, we store the triangular factors and use triangular solves. The cubic complexity of computing the initial factorization then dominates the quadratic complexity of the solve step. This approach is numerically more stable and has the same complexity as computing the explicit inverse and performing matrix-vector multiplication.

Second, if we solve for $x_1^{k+1}$ in equation \eqref{eq:eq_system}, instead of $v^k$, we need to invert
the Schur complement $S' = A^\top A + I$ corresponding to the lower-right submatrix. This would be advantageous if $A$ is taller than it is wide. In our case, it is the other way around since the number of columns in $A$ (determined by the number of dose points) is typically much larger than the number of rows (determined by the number of isocenters times the number of degrees of freedom per isocenter).\looseness=-1

Unlike the simplex algorithm, which finds the exact optimum (up to round-off error) after a finite number of steps, ADMM converges gradually to the optimal solution. Hence, for the latter, we either need to fix the number of iterations or use a suitable stopping criterion. In addition, the practical performance can be improved by tuning the step size $\rho$, and preconditioning the constraint matrix $A$.

\subsection{Two-pass optimization}
The final component of our method addresses the need to control dose fall-off, and thereby the dose exposure to healthy brain tissue, while maintaining target coverage and respecting dose constraints to  OARs. The approach, which emerged when bringing the method of Sjölund et al.\cite{sjolund2019linear} into clinical practice, incorporates adaptive selection of dose points that contribute to the gradient-index term in the cost function. This selection is updated iteratively to better reflect the regions where dose shaping is most needed. In essence, we aim to focus the optimization effort on those regions where the dose is close to a specified low-dose threshold, $D_\text{LD}$. Here, as in the clinical software, two iterations are performed and thus, for each set of (primal) weights, two optimization problems of the same form as in \eqref{eq:dual_problem} are solved. We refer to this procedure as a \emph{two-pass optimization}. It bears some similarity to other methods for incorporating DVH constraints\cite{zarepisheh2013moment,fu2019convex}, but is specifically tailored to the parallel setting. 

\subsubsection{Definition of the low-dose volume}
What differs between the passes is the selection of low-dose points, i.e., those points sampled from the low-dose volume and penalized according to the low-dose term of the objective function. 
In the first pass, the low-dose points are sampled from a low-dose volume calculated from a geometric expansion of the target, selected to encompass approximately the volume where the dose is expected to be close to the low-dose threshold $D_\text{LD}$. In the second pass, the low-dose points are instead sampled from a low-dose volume calculated from the dose distribution corresponding to the first-pass optimization. This is achieved by selecting a dose range around the low-dose  threshold , $[D_\text{LD}, D_\text{LD} + \Delta)$, and letting the low-dose volume be the portion of patient volume where the first-pass dose distribution is within this range. Although this can be viewed as an iterative procedure, we restrict our attention to just the first two iterations, as they correspond to the clinical implementation. Extensions of the approach to multiple low-dose thresholds, as well as handling multiple targets and OARs, are described in Appendix~\ref{app:dual_lp}. 

\subsubsection{Non-redundant sampling of dose points}\label{sec:sampling}
In the first pass, we sample dose points independently of the objective function weights, which means the parallel approach from Section~\ref{sec:admm} is directly applicable. However, in the second pass, the sampling volumes are defined by the solutions to the first-pass problem, which differ depending on the weights. This dependency makes it non-trivial to solve the second-pass problem in parallel.

One way around this is to
sample from the union of all low-dose volumes and, for each weight vector, zeroing out elements in the upper bound ($u$ in Appendix~\ref{app:dual_lp}) corresponding to dose points outside the low-dose volume for the selected weight vector. However, this does not allow sampling low-dose points with different densities for different weight vectors, which is desirable to ensure a constant, minimum, or maximum number of low-dose points.
Next, we describe a method that ensures overlapping low-dose volumes are represented by a single set of low-dose points, with maximal reuse of points whilst retaining (approximately) the desired sampling density for every weight vector.

Let $V_j^\cup$ and $p_j^\cup$ be the union of all low-dose volumes and points. As shown in Algorithm~\ref{alg:sampling}, 
we start with $V_0^\cup = \emptyset$ and $p_0^\cup = \emptyset$ and iteratively sample points $p_j$ from volumes $V_j$ with probability $\omega_j = N_{\text{LD},j}/|V_j|$, where the volumes $V_j$ are ordered according to decreasing sampling density $\omega_1 \geq \omega_2 \geq \ldots \geq \omega_m$.
\begin{algorithm}[t]
\caption{Overlapping low-dose point sampling}
\begin{algorithmic}[1]
\State Initialize $V_0^\cup \gets \emptyset$, $p_0^\cup \gets \emptyset$
\For{$j = 1$ to $m$}
    \State $\tilde{p}_j^\cup \gets$ randomly select $\omega_j |V_{j-1}^\cup|$ points from $p_{j-1}^\cup$
    \State $p_j^{\text{in}} \gets \{p : p \in \tilde{p}_j^\cup \cap V_j \}$
    \State $p_j^{\text{out}} \gets$ sample $\omega_j |V_j \setminus V_{j-1}^\cup|$ points from $V_j \setminus V_{j-1}^\cup$
    \State $p_j \gets p_j^{\text{in}} \cup p_j^{\text{out}}$
    \State $p_j^\cup \gets \tilde{p}_j^\cup \cup p_j^{\text{out}}$
    \State $V_j^\cup \gets V_j \cup V_{j-1}^\cup$
\EndFor
\State \Return $P = \bigcup_j p_j$
\end{algorithmic} \label{alg:sampling}
\end{algorithm}
The union of all low-dose points required to calculate $A$ is then given by $P = \bigcup_j p_j$.

We note that since $p_j^{\text{in}}$ depends on the random distribution of points from previous steps, the number of points in this set will follow a binomial distribution and thus equal the desired $\omega_j |V_j \cap V_{j-1}^\cup|$ only in expectation. The standard deviation of the distribution is given by $\sqrt{|\tilde{p}_j^\cup|r(1-r)}$, where $r=|V_j \cap V_{j-1}^\cup|/|V_{j-1}^\cup|$, the fraction of $V_{j-1}^\cup$ intersected by $V_j$. Since $|\tilde{p}_j^\cup|r = \omega_j |V_j \cap V_{j-1}^\cup| \leq \omega_j |V_j|$, the standard deviation relative to the total desired number of points in $V_j$ is given by
\begin{equation}
\frac{\sqrt{|\tilde{p}_j^\cup|r(1-r)}}{\omega_j |V_j|} \leq \frac{\sqrt{\omega_j |V_j|(1-r)}}{\omega_j |V_j|} = \frac{\sqrt{(1-r)}}{\sqrt{\omega_j |V_j|}} < \frac{1}{\sqrt{\omega_j |V_j|}}.
\end{equation}
Hence, the worst-case relative standard deviation is inversely proportional to the square root of the desired number of points in volume $V_j$. Selecting a minimum number of points of a few hundred hence limits the worst-case standard deviation to a few percent.

\section{Results}

\subsection{Implementation and hardware}
We evaluate a CPU and a GPU version of the proposed ADMM algorithm. Both are implemented in Matlab (The MathWorks Inc., Natick, Massachusetts, USA) with the only difference being that the GPU code uses the Parallel Computing Toolbox to run the ADMM iterations on a GPU.
The baseline method is to solve problems sequentially using Matlab's built-in, single-threaded solver \texttt{linprog}, using the dual simplex algorithm called ``Dual Simplex Legacy" in version 2024b\cite{matlab}. This algorithm was chosen since, on average, it was seen to be faster than the ``Dual Simplex HiGHS" algorithm on the validation cases.

Matlab is a high-level language typically associated with less efficient use of computational resources than lower-level languages. However, the ADMM iterations rely almost exclusively on linear algebra operations, for which Matlab uses highly efficient low-level subroutines. Hence, the performance penalty is expected to be modest. However, as of version 2024b, Matlab does not support caching objects of the built-in, low-level matrix decomposition class on the GPU. To avoid recalculating the decomposition in every iteration of the algorithm, we therefore use the explicit inverse of the Schur complement $S$ (and do so also on the CPU). In terms of computations per iteration, this should be similar to an efficient decomposition method, but could worsen the numerical stability of the algorithm. However, our experimental results suggest this was not a problem for our particular cases.
 
To make the most of the GPU parallelism, we rely on single-precision floating-point arithmetic in the ADMM iterations (but perform the explicit inversion of $S$ in double precision on the CPU before casting to single precision and performing the iterations on the GPU). We do the same when running the ADMM iterations on the CPU.

To measure execution time, we first make a call to the function once without timing it to exclude the overhead of the just-in-time compiler (with the expectation that Matlab caches the compiled functions for subsequent runs). After this, the function is executed several times and the median execution time is reported. For a fair comparison, we include memory transfer times to and from the GPU. Apart from this, we measure only the time required to solve the optimization problems; no effort was made to speed up intermediary steps like the dose calculation after each optimization pass, since this was considered beyond the scope of the current investigation. However, as detailed in the Discussion, this is not expected to impact on the applicability of the method.

Experiments were performed on a workstation running Windows 11 equipped with two Intel Xenon Gold 6248R CPUs, each with 24 physical cores, a total of 256 GB of CPU RAM, and an Nvidia RTX A6000 GPU with 10752 FP32 cores and 48 GB of GPU RAM.

\subsection{Test cases and metrics}\label{sec:data}
The test cases consisted of anonymized patient data from a total of 20 treatments selected from three common Gamma Knife indications: meningioma, vestibular schwannoma (acoustic neuroma), and multiple metastases. These intentionally span a range of target sizes, with an emphasis on larger cases and those including OARs, which are generally more difficult and time-consuming to optimize. The cases included clinical target and OAR volumes and a skull mask, as well as isocenter positions and dose rate kernels, all exported from GammaPlan. A summary of the test cases is provided in Table~\ref{tab:case_characteristics}. To evaluate different preconditioning techniques and values of $\rho$, we used a separate validation set consisting of ten other cases with the same three indications as the test set.

To compare the different optimization methods, we use the clinical metrics coverage, selectivity, gradient index and beam-on time. For single-target cases, these are defined as in Section~\ref{sec:formulation}. The beam-on time is computed assuming a calibration dose rate of 3 Gy/minute. The definitions of coverage and beam-on time generalize readily to multi-target cases, as does the definition of selectivity, provided that the volumes receiving the prescription dose for each target do not overlap. A global gradient index, however, can only be defined if all prescription doses are equal. When this is not the case, we leave out this metric.\looseness=-1

In addition to the clinical metrics, we compare the computation times required to perform the first and second optimization passes for multiple weight vectors for the test cases. For each case,
the weight vectors $(w_\text{T}^j, w_\text{R}^j, w_\text{LD}^j, w_\text{BOT}^j)$ were generated corresponding to unique combinations of the ‘low-dose’ and ‘beam-on time’ sliders, $\{s_\text{LD},s_\text{BOT}\}$, in the clinical software.

\subsection{Performance tuning}\label{sec:tuning}
In the current work, we limit the investigation of preconditioning to scaling rows and columns of $A$ (together with the corresponding scalings of $c$, $b$, and $u$). We tuned $\rho$ by performing a grid search  over $\rho \in \{0.1 / 2^n\}$ with $n = 0, 1, \ldots, 15$ on the validation data described in Section~\ref{sec:data}. For each value, we performed a fixed number of ADMM iterations on the first-pass problem and then compared the objective function terms of the primal problem with those obtained for the exact same problem solved by the simplex method. 

As seen in Table \ref{tab:case_characteristics}, the number of elements and the shape of the constraint matrix $A$ vary considerably between patient cases due to differences in the number of targets, total target volume and surface area, number of OARs, and number of isocenters. Hence, without preconditioning, the magnitude of the elements of the Schur complement $S$, and thereby the behavior of the first step of the ADMM algorithm in Equation~\eqref{eq:x-step}, is highly case dependent. To alleviate this, we first normalize the rows of $A$, causing the diagonal elements of $S$ to equal two and all remaining elements to lie on the interval $[-1, 1]$, which was seen to make the performance on the validation cases more consistent.

\begin{table}[tb!]
\begin{center}
\vspace{2mm}
\renewcommand{\arraystretch}{1.05}
\captionv{15}{Case and problem characteristics for the clinical test set}{Summary of attributes, geometric complexity and first-pass constraint matrix sizes for the 20 clinical test cases. The naming of the cases reflects the indications: meningioma (ME), multiple metastases (MM) and vestibular schwannoma (VS).}
\label{tab:case_characteristics}
\vspace*{2ex}
\begin{tabular}{l|ccccc}
\hline
\textbf{Case ID} & \makecell{\textbf{Number of}\\\textbf{targets}} & \makecell{\textbf{Total target}\\\textbf{volume [mm\textsuperscript{3}]}} & \makecell{\textbf{Number of}\\\textbf{OARs}} & \makecell{\textbf{Number of}\\\textbf{isocenters}} & \makecell{\textbf{First-pass}\\\textbf{problem size}} \\
\hline
ME01 &  1 & 1943  & 1 & 53  & 1325 $\times$ 4547  \\
ME02 &  1 & 7573  & 1 & 96  & 2400 $\times$ 9486  \\
ME03 &  1 & 7618  & 1 & 79  & 1975 $\times$ 9498  \\
ME04 &  1 & 11004 & 1 & 117 & 2925 $\times$ 10096 \\
ME05 &  1 & 11068 & 1 & 202 & 5050 $\times$ 11589 \\
ME06 &  1 & 14229 & 2 & 95  & 2375 $\times$ 15413 \\
ME07 &  1 & 15155 & 2 & 68  & 1700 $\times$ 16857 \\
ME08 &  1 & 23839 & 1 & 177 & 4425 $\times$ 18991 \\
ME09 &  1 & 26749 & 1 & 105 & 2625 $\times$ 19468 \\
ME10 &  1 & 37270 & 1 & 214 & 5350 $\times$ 26537 \\[1ex]
MM01 &  4 & 9552  & 0 & 62  & 1550 $\times$ 16295 \\
MM02 &  2 & 14042 & 1 & 58  & 1450 $\times$ 20482 \\
MM03 &  3 & 28546 & 0 & 117 & 2925 $\times$ 31090 \\
MM04 &  9 & 29018 & 0 & 145 & 3625 $\times$ 34925 \\
MM05 &  2 & 52400 & 0 & 113 & 2825 $\times$ 40718 \\[1ex]

VS01 &  1 & 663   & 2 & 17  & 425 $\times$ 4046   \\
VS02 &  1 & 4349  & 2 & 25  & 625 $\times$ 4870   \\
VS03 &  1 & 5850  & 2 & 32  & 800 $\times$ 9564   \\
VS04 &  1 & 8274  & 2 & 21  & 525 $\times$ 7645   \\
VS05 &  1 & 12629 & 1 & 49  & 1225 $\times$ 10179 \\
\hline
\end{tabular}
\end{center}
\end{table}

For each of the validation cases, after normalizing the rows of $A$, 2000 ADMM iterations were run on the first-pass problem, for each of the 16 values of $\rho$ and 121 weight vectors corresponding to all combinations of low-dose and BOT slider settings between 0.0 and 1.0 in increments of 0.1. The values of the terms of the primal cost function were then compared with those of the true solutions, obtained with the simplex solver, for the exact same problems. We observed that a good agreement between the total cost and the BOT term (which is proportional to the actual BOT of the treatment) of the cost function indicated an acceptable quality of the solution. We found that the optimal values of $\rho$ with respect to these two metrics generally decrease with increasing BOT penalization (larger $w_\text{BOT}$) but were less impacted by the values of $w_S$ and $w_G$. We thus constructed the rule
\begin{equation}
    \rho = 2.5 \times 10^{-3} \sqrt{\frac{w_{0,\text{BOT}}}{w_\text{BOT}}},\label{eq:rho_rule}
\end{equation}
where $w_{0,\text{BOT}}$ is the lowest allowed value of the BOT weight. This produced close to the best agreement with the true values of both the total cost function and its BOT term for all validation cases and weight vectors. Since $\rho$ enters the ADMM algorithm only through the iterates, we can choose a unique value for $\rho$ for each weight combination, even when solving multiple problem instances at once. Hence, the $w_\text{BOT}$-dependent value of $\rho$ in Equation~\eqref{eq:rho_rule} was used in all subsequent experiments.

Although normalizing the rows of $A$ and adopting the value of $\rho$ above produced good general agreement in the primal cost function for all validation cases, the agreement in the BOT term was somewhat worse for the largest case, with relative differences up to 12\% for a few weight vectors. For large cases, the number of dose points can become large relative to the number of isocenters. Since the number of columns of $A$ related to the BOT 
is proportional to the number of isocenter positions, the relative importance of the BOT decreases as the number of dose points increases. To counteract this, we investigate scaling the BOT columns of $A$ (after normalizing the rows) by a factor $\beta$ proportional to the number of dose points (not including OAR dose points since these do not contribute to the cost function of the primal problem). In the notation of appendix~\ref{app:dual_lp}, choosing
\begin{equation}
    \beta = \frac{N_\text{TS} + N_\text{TI} + N_\text{R} + N_{1,\text{LD}} + N_{2,\text{LD}}}{2000},
\end{equation}
produces values of $\beta$ ranging from about 1 (i.e., no scaling) for smaller cases up to about 20 for the largest cases considered here.

The above choice of $\rho$ and scalings of $A$ result in the overall objective function value agreeing to within 3\% for all validation cases and all weight vectors after 2000 iterations, with a mean absolute deviation of 0.11\%. For the BOT term, the agreement was within 5\% for all cases and weight vectors, with a mean absolute value of 0.24\%. Increasing the number of iterations to 3000 further improved the agreement to within about 1\% and 2\%, respectively, for the total objective and BOT term.

\subsection{Quality of ADMM solutions}
To evaluate the quality of the ADMM solutions, we compare the clinical metrics after the second-pass optimization to the corresponding metrics for the baseline simplex solver. To quantify the noise in clinical metrics caused by the random sampling of dose points, the optimizations for each of the test cases were rerun 20 times, with a new random selection of dose points for each run. To span the range of the slider space without making the number of optimizations infeasibly large, each test case was optimized for nine weight vectors corresponding to all combinations of low-dose and BOT slider settings in $\{0.0, 0.5, 1.0\}$. Hence, in total, $20 \times 20 \times 9 = 3600$ sets of metrics were obtained for each solver.

For the simplex solver, optimization was performed sequentially with unique low-dose points sampled for each second-pass optimization. For the largest cases and some weight vectors, some unfortunate samplings of dose points resulted in problem instances where the simplex solver took more than an order of magnitude longer to finish than for others. The normalization of the rows of $A$ alleviated this problem and somewhat reduced the average solution time also for other cases, and was hence used. Additionally, the simplex solver was run with the `Preprocess' option enabled, as this proved beneficial in some cases. For the ADMM solver, we applied the performance tuning from the previous section and, in both optimization passes, ran 3000 iterations in parallel for all weight vectors. The second-pass low-dose points were sampled as described in Section~\ref{sec:sampling}.

\begin{figure}[tb!]
\begin{center}
\includegraphics[trim={1.5cm 0 1.5cm 0},clip,width=\textwidth]{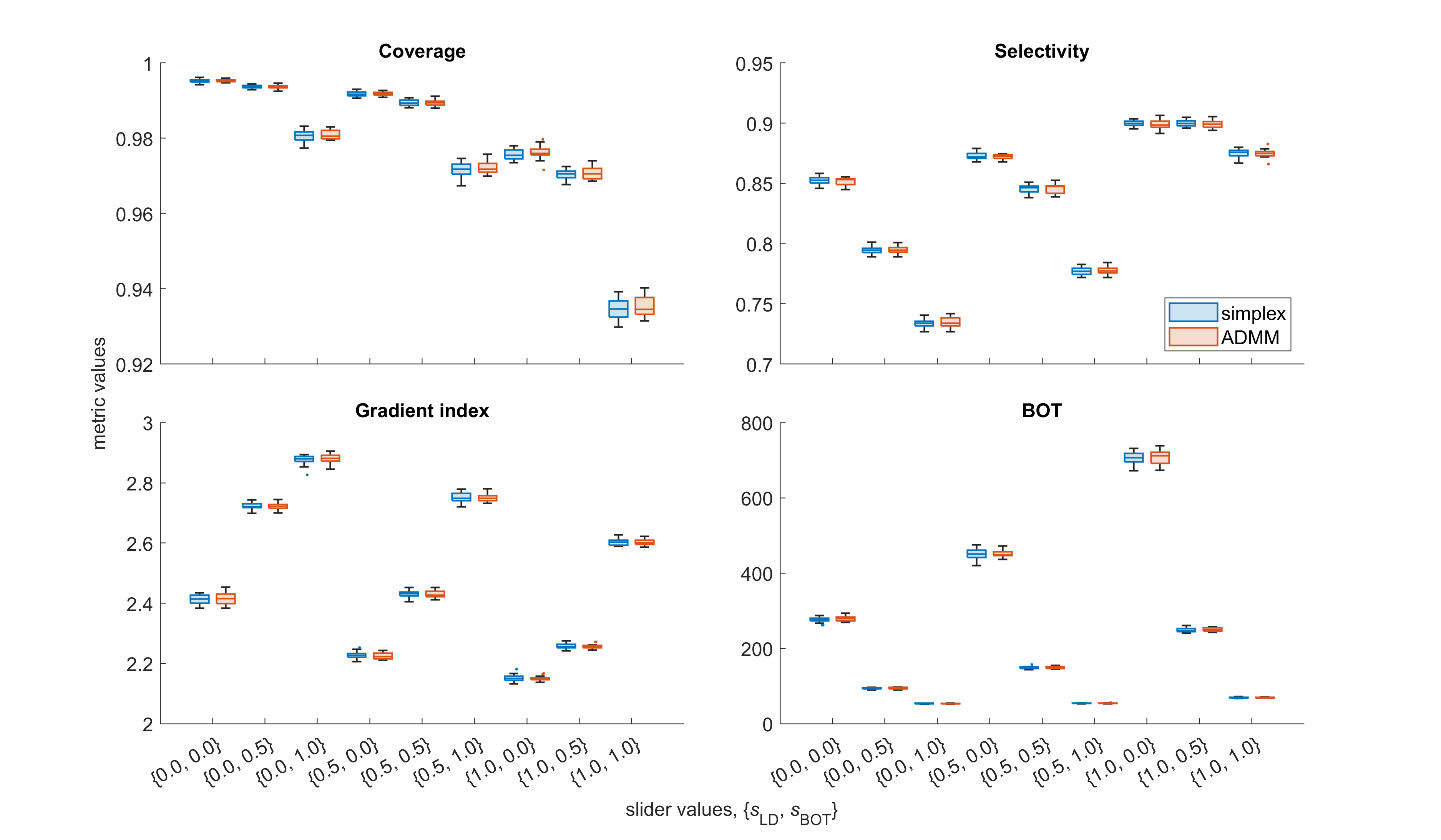}
\caption{Distributions of clinical metrics for case ME10 after the second optimization pass. Each box plot corresponds to 20 reruns of the same weight vector. Lines inside a boxes represent the median values, top and bottom edges the quartiles, whiskers the non-outlier maxima and minima, and dots outliers. Outliers are defined as points further away than 1.5 inter-quartile ranges from the top and bottom of a box.}
\label{fig:metrics_ME10}
\end{center}
\end{figure}

Figure~\ref{fig:metrics_ME10} shows the spread in clinical metrics after the second optimization pass over 20 reruns for the simplex and ADMM solvers for the largest single-target test case, ME10. For this case, as well as the remaining 19 test cases, the agreement between the distributions of metrics is good for all weight vectors. More specifically, we quantify this agreement by considering the distributions of relative differences between the ADMM metrics for each test case and weight vector with the mean of the corresponding simplex metrics. The mean and standard deviation over these 3600 sets of normalized ADMM metrics are:
$0.0 \pm 0.2\%$ for coverage,
$0.0 \pm 1.6\%$ for selectivity,
$0.1 \pm 0.8\%$ for gradient index, and
$0.1 \pm 3.0\%$ for BOT.
The corresponding values for the distributions of normalized simplex metrics were:
$0.0 \pm 0.2\%$ for coverage,
$0.0 \pm 1.7\%$ for selectivity,
$0.0 \pm 0.7\%$ for gradient index, and
$0.0 \pm 2.8\%$ for BOT.
(By definition, the normalization makes the mean values of the simplex metrics zero.)

\begin{figure}[tb!]
\begin{center}
\includegraphics[trim={1.5cm 0 1.5cm 0},clip,width=\textwidth]{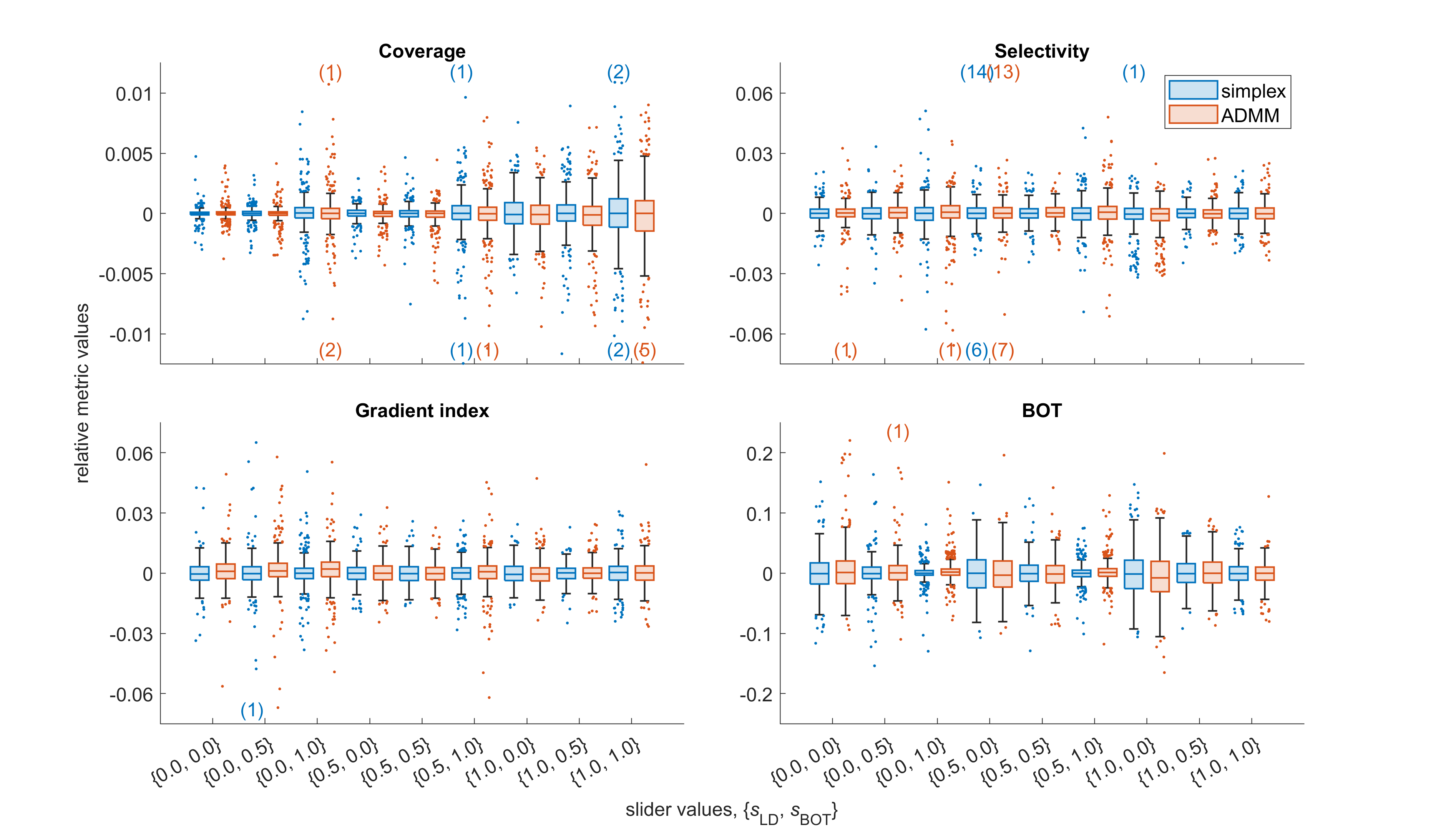}
\caption{Distributions of clinical metrics after the second optimization pass, normalized to the mean simplex value for each case and weight vector, grouped by weight vector. Each box plot represents 400 optimizations, corresponding to 20 reruns of 20 different cases, except for metrics that do not apply to all cases (see text). Numbers in parentheses indicate outliers above or below the $y$-axis limits. Other box properties are the same as in Figure~\ref{fig:metrics_ME10}.}
\label{fig:normalized_metrics_per_weight_set}
\end{center}
\end{figure}

\begin{figure}[ht!]
\begin{center}
\includegraphics[trim={1.5cm 1.5cm 1.5cm 1.5cm},clip,width=\textwidth]{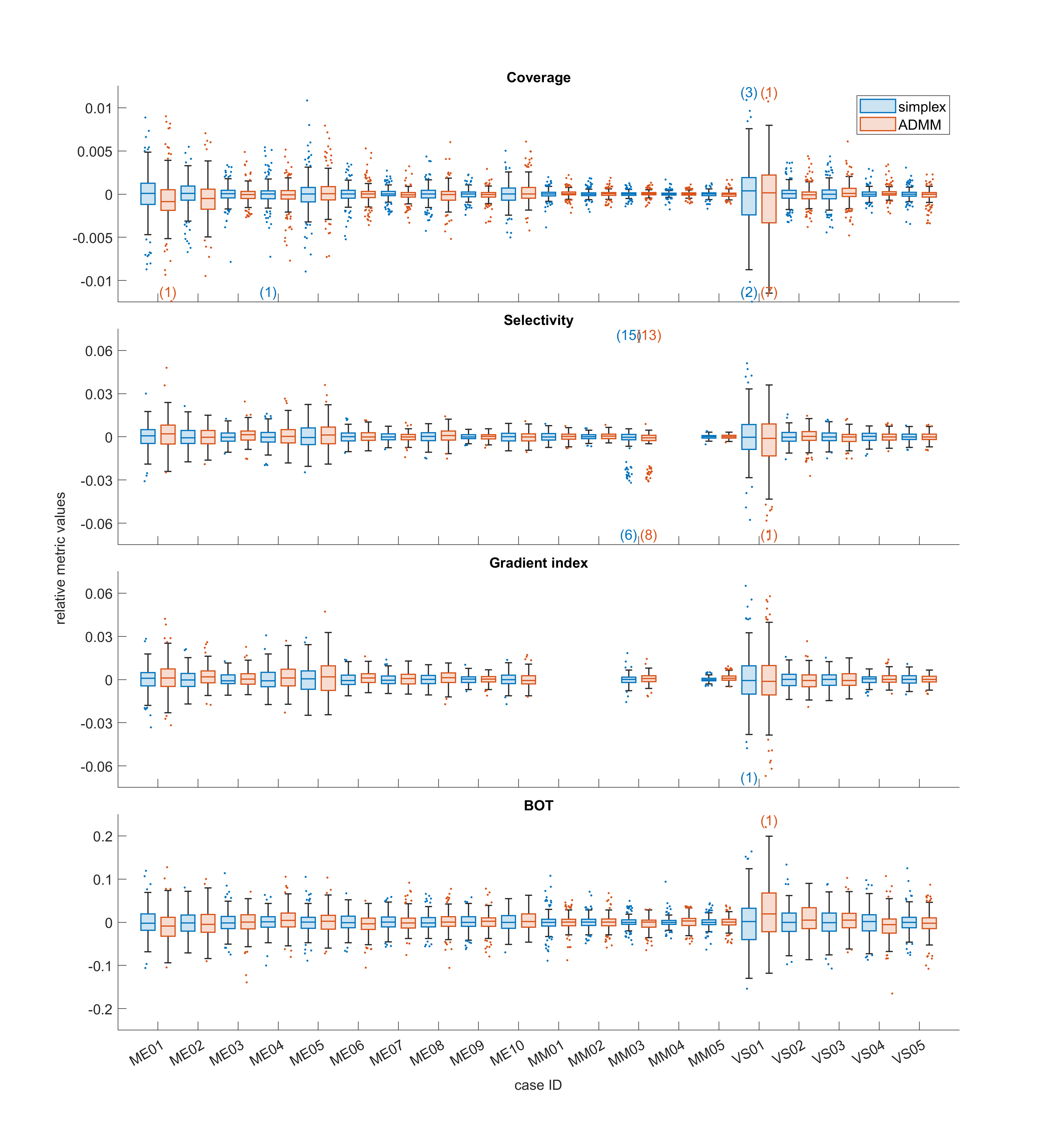}
\caption{Distributions of clinical metrics after the second optimization pass, normalized to the mean simplex value for each case and weight vector, grouped by case. Each box plot represents 180 optimizations, corresponding to 20 reruns of 9 different weight vectors. Missing boxes indicate metrics not applicable to a given case, as described in the text. For box properties, see Figures~\ref{fig:metrics_ME10} and \ref{fig:normalized_metrics_per_weight_set}.}
\label{fig:normalized_metrics_per_case}
\end{center}
\end{figure}

To further visualize these results, Figures~\ref{fig:normalized_metrics_per_weight_set} and \ref{fig:normalized_metrics_per_case} show the distributions of normalized simplex and ADMM metrics, split per weight vector and per test case, respectively. Missing bars in Figure~\ref{fig:normalized_metrics_per_case} indicate that the metric is not applicable for the corresponding case. In particular, the selectivity becomes nonsensical for MM04 due to the proximity of some of the targets, and the gradient index is not well defined for MM03, MM04 and MM05 due to different prescription doses to different targets, as discussed above. Hence, for these particular cases, these metrics are not included in the distributions shown in Figure~\ref{fig:normalized_metrics_per_weight_set}, nor in the analysis in general.

For two of the 180 combinations of test cases and weight vectors, a systematic difference (i.e., where the difference in mean value between the ADMM and simplex metrics was much larger than the standard deviation in the simplex metric) in one of the metrics was identified. For MM04, the case with the most individual targets, the mean ADMM BOT was systematically higher than the mean simplex BOT for the weight vector corresponding to $\{s_\text{LD}=0, s_\text{BOT}=1\}$. In absolute terms, the difference in the mean values was just under 2 minutes (58.9 versus 57.0 minutes), which we deem to be of little, if any, clinical relevance. Similarly, the mean ADMM gradient index for MM05, the case with the largest total target volume, was systematically higher than that of the simplex gradient index for the same combination $\{s_\text{LD}=0, s_\text{BOT}=1\}$. In this case, the absolute difference was less than 0.02 (2.72 versus 2.70), again deemed to be of little or no relevance.

\subsection{Timing results}\label{sec:timing}
\begin{figure}[htb]
\begin{center}
\includegraphics[trim={1.5cm 0.5cm 1.5cm 1cm},clip,width=\textwidth]{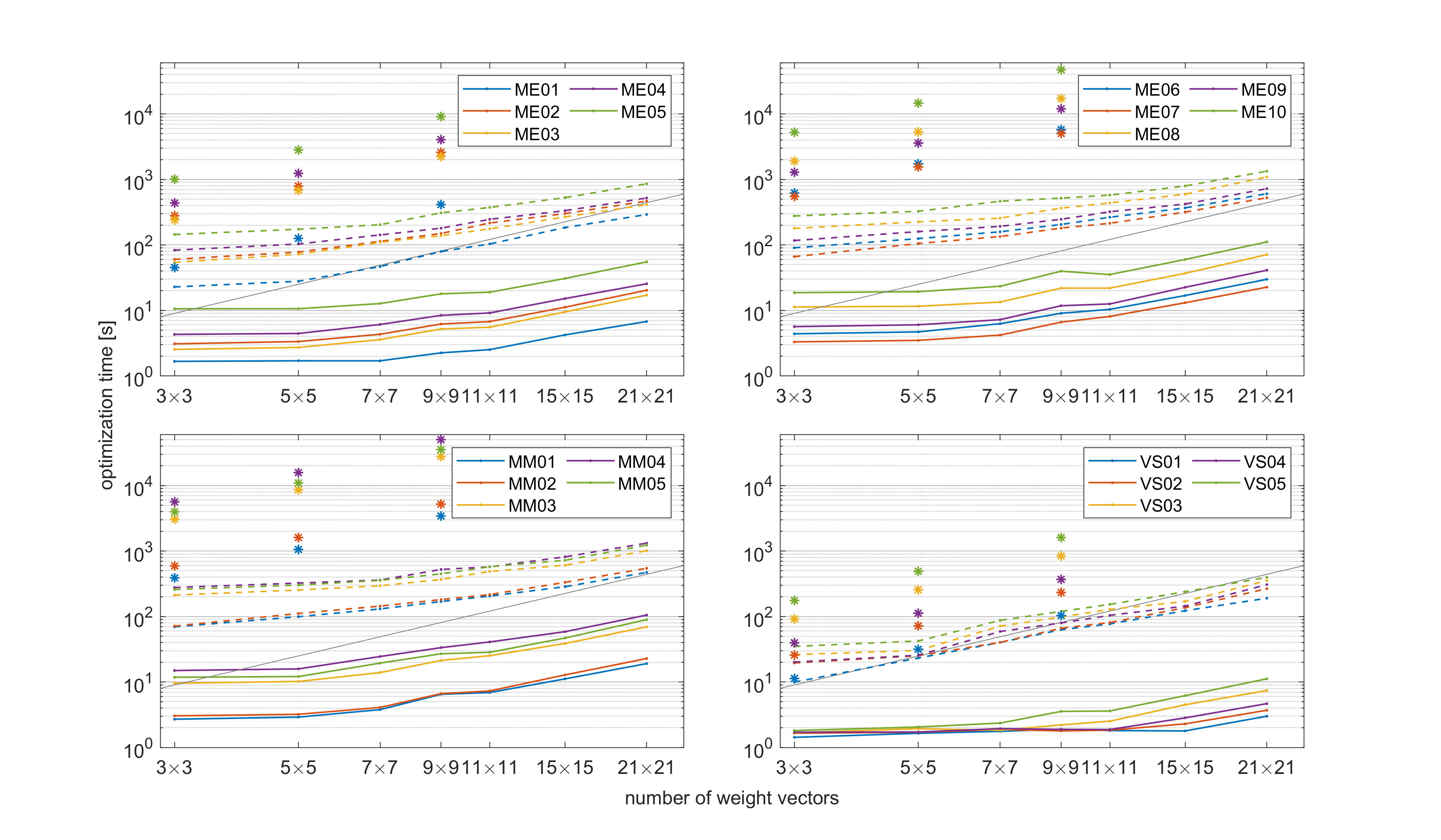}
\caption{Log-log plots of ADMM optimization times in seconds for two passes of 3000 iterations each, for different numbers of weight vectors on the GPU (solid line) and CPU (broken line). Two-pass simplex times for three sets of weight vectors are shown as asterisks. The slope of the gray diagonal lines indicates linear growth. The top-left panel shows small- and medium-sized meningioma cases (up to 11 cm\textsuperscript{3}) and the top-right panel large-sized meningioma cases (above 14 cm\textsuperscript{3}). The left and right panels at the bottom show multiple metastasis and vestibular schwannoma cases, respectively. Results are the sums of the median times for the first and second optimization passes over ten runs (five runs for simplex).
}
\label{fig:timing_results}
\end{center}
\end{figure}

Timing results for running two optimization passes of 3000 ADMM iterations each, for all test cases on both CPU and GPU, are shown in Figure~\ref{fig:timing_results}. Results were obtained for between nine and 441 weight vectors, corresponding to between $3\times3$ and $21\times21$ slider values. For reference, the figure also shows single-threaded simplex timings for comparison, for between nine and 81 weight vectors.  For nine weight vectors, corresponding to the $3\times3$ slider values from the previous section, the simplex solver required between 12 and 5700 seconds, depending on the test case. The corresponding ranges for the CPU and GPU implementation of the ADMM algorithm were 10--280 seconds and 1.4--19 seconds, respectively. For 81 weight vectors, the timing ranges were 100--51000, 63--520, and 1.8--40 seconds for simplex, ADMM on CPU, and ADMM on GPU, respectively.

When the number of weight vectors is small, the ADMM run times grow sub-linearly with the number of weight vectors. This is expected due to the constant overhead of the ADMM method (mainly inverting the Schur complement) and because the number of operations in each ADMM iteration is not large enough to fully utilize the hardware. As a result, for many cases, a five-fold increase in the number of weight vectors, from $3\times3$ to $7\times7$, comes with a small time penalty. Similarly, optimization times for $9\times9$ and $11\times11$ weight vectors were very similar on the GPU. For the largest sets of weight vectors, however, we see the expected linear growth in optimization times, where optimizing 441 instances in parallel required between 3.0 and 110 seconds for the different test cases.

\begin{figure}[htb]
\begin{center}
\includegraphics[trim={1.5cm 0.5cm 1.5cm 1cm},clip,width=\textwidth]{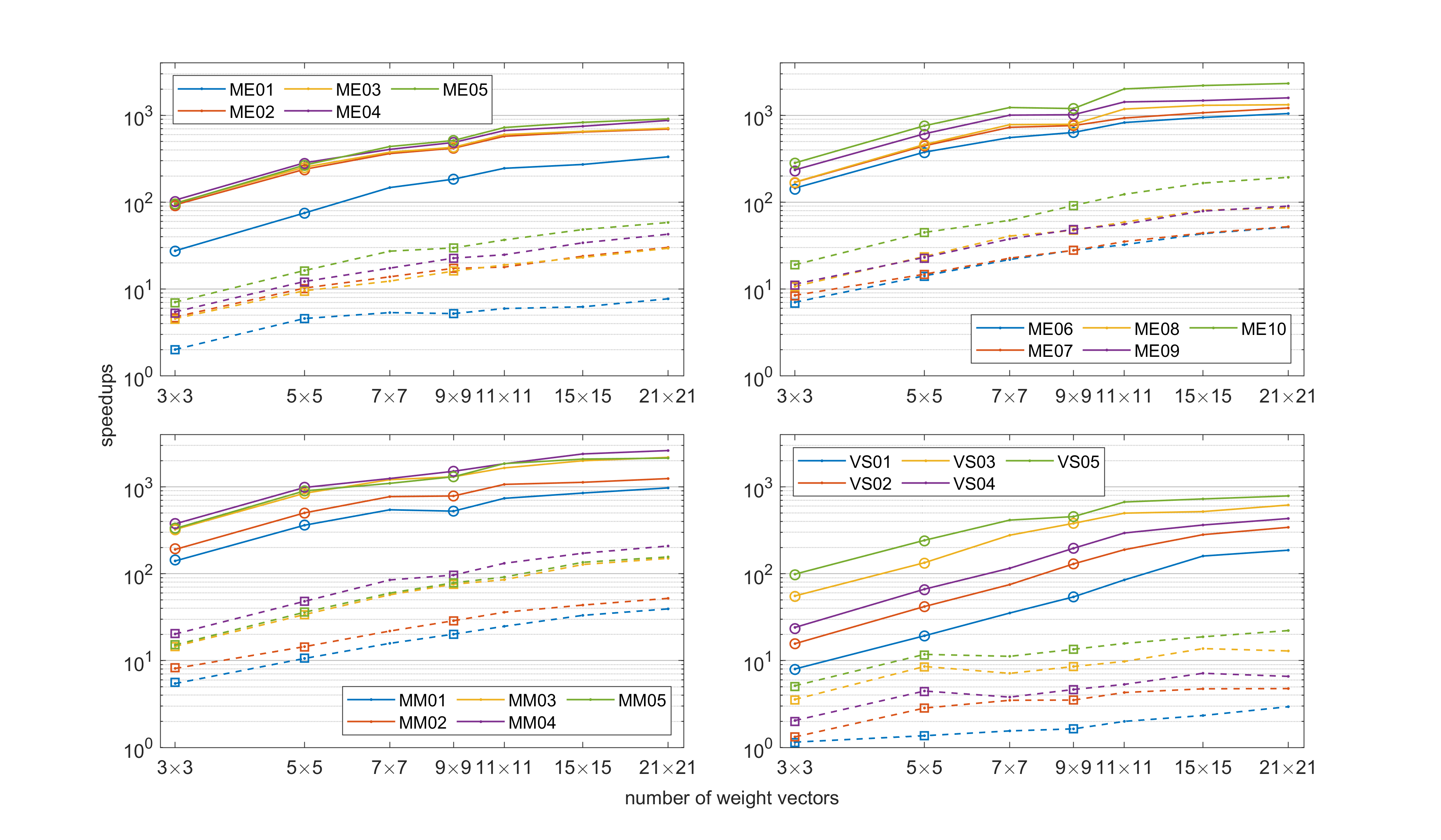}
\caption{Log-log plots of speedups when using the ADMM method on GPU (rings) and CPU (squares) compared to the single-threaded simplex method. Solid and broken lines represent estimated speedups for GPU and CPU, respectively, based on linear scaling of the simplex times for $9\times9$ weight vectors (see text). The top-left panel shows small- and medium-sized meningioma cases (up to 11 cm\textsuperscript{3}) and the top-right panel large-sized meningioma cases (above 14 cm\textsuperscript{3}). The left and right panels at the bottom show multiple metastasis and vestibular schwannoma cases, respectively.
}
\label{fig:speedup_results}
\end{center}
\end{figure}

From the above results optimizing for nine weight vectors, the ADMM solver on CPU achieves a speedup between 1.2 and 20 times compared to the single-threaded simplex solver. The corresponding speedup when running the ADMM solver on GPU for the same number of weight vectors is between 8.2 and 380 times. Moving to 81 weight vectors, speedups were 1.6--97 times and 54--1500 times for ADMM on the CPU and GPU, respectively. The large spans seen in the speedups can again be attributed to hardware utilization, with larger problems resulting in higher relative efficiency of the ADMM algorithm. Hence, importantly, the largest speedups are seen for the largest cases, both on CPU and GPU, where the optimization times are the longest. The speedups for individual cases are shown in Figure~\ref{fig:speedup_results}. 

As opposed to the ADMM method, the reference simplex method solves each problem corresponding to a weight vector sequentially, and the simplex times are thus expected to be linear in the number of weight vectors, as can be verified in Figure~\ref{fig:timing_results}. This allows us to estimate the simplex run times for a larger number of weight vectors than would be convenient to optimize. For reference, the ADMM speedups when optimizing for up to 441 weight vectors in parallel, compared to the corresponding estimated simplex run times, are also shown in Figure~\ref{fig:speedup_results}. 

\section{Discussion}

\paragraph{Relevance}
Our experiments show that, with the basic tuning of the algorithm described in Section~\ref{sec:tuning}, a fixed number of 3000 iterations per optimization pass is sufficient for our ADMM algorithm to replicate the simplex plan quality to within clinical relevance in all test cases. 
Importantly, we demonstrate hundredfold speed-ups as compared to sequential simplex runs when solving more than a hundred weights in parallel.
In particular, we can generate 441 Pareto-optimal plans, corresponding to $21\times 21$ slider settings in the clinical software, in less than two minutes --- even for the largest test cases. We consider this fast enough to bring multicriteria optimization well within the time frame of the interactive clinical workflows typically associated with radiosurgery planning. We further expect that our methodology can be extended to multicriteria optimization for other convex fluence map optimization problems, e.g. in intensity-modulated radiotherapy, by approximating nonlinear functions as piecewise linear ones\cite{romeijn2003novel}.

\paragraph{Timing comparisons} One important caveat of the timing comparisons between ADMM and simplex, summarized in Figure~\ref{fig:timing_results}, is that the simplex method runs on a single thread while ADMM leverages parallelism in both the CPU and GPU cases. Since the two CPUs have 48 cores in total, it is reasonable to ask whether one can run one simplex algorithm on each. The short answer is, however, no --- the simplex method uses too much memory. For the largest cases, which is where ADMM was the most beneficial, we could only run four simplex instances in parallel despite having 256 GB of CPU RAM. For the implementation used, the memory usage was further seen to be instance dependent and difficult to predict, so that for some unfortunate weight vectors, the solver ran out of memory with only three parallel instances.

As mentioned, the built-in preprocessing was beneficial for some validation cases and was hence used with the simplex solver. A natural improvement would be to run the preprocessing once for all instances of a case, but since this was not possible in the Matlab solver, it was not explored. However, especially for larger cases, the time spent on the preprocessing was small compared to the total optimization time, which means that, at least for the tested implementation, the benefits of this improvement would be modest.

\paragraph{Additional steps}
In addition to the optimization itself, plan optimization involves some additional steps. First, the isocenter placement, which with the presented method is independent of the objective weighting. Hence, since the placement is done once regardless of the number of weight vectors, and the calculation itself is fast, this step has no noticeable effect on the overall optimization time.

Second, we perform a dose calculation after each of the two optimization passes. Although we consider this beyond the scope of the current investigation, we note that the dose calculation for a given weight vector is the sum of a number of precalculated dose-rate kernels, weighted by the sector times from the corresponding optimization. Since the dose calculation for each weight vector uses (a subset of) the same dose-rate kernels, dose calculation for multiple weight vectors should lend itself well to parallel implementation. In addition, limiting the number of dose points in the calculation to those in the vicinity of the target necessary for calculating the second-pass low-dose points (after the first pass) and relevant clinical metrics (after the second pass) should allow further speeding up the dose calculation. Hence, we expect that the inclusion of dose calculation times should not affect the applicability of the presented method.

Finally, the sequencing step, where the sector times from the optimization step are converted into a deliverable treatment plan, has not been considered in this work. This we justify by the fact that, in contrast to conventional radiotherapy --- where the sequencing can be challenging and accounting for machine limitations can have a non-negligible impact on plan quality\cite{shepard2002direct} --- this step is virtually trivial, and has a minor dosimetric impact, for Gamma Knife treatments. (In fact, if we disregard pruning of shots that are too short to be deliverable and the time required to move between isocenters, the sequenced plan parameters are guaranteed to be identical to the ones produced by the optimizer.) Hence, we imagine the interactive Pareto navigation to be based directly on the plans from the second optimization pass.\looseness=-1

\section{Conclusions}
Running the presented adaptation of the ADMM algorithm on GPU, it is possible to optimize several hundreds of radiosurgery treatment plans, corresponding to different clinical trade-offs, in a couple of minutes. This is true also for large targets and without compromising plan quality relative to plans optimized by the method in current clinical use. Integrated into a treatment planning system, we foresee that such a tool would allow multi-criteria optimization followed by interruption-free plan navigation within the strict time frame set by the clinical workflow.

\section*{Acknowledgments}
We would like to thank Dr Josef Novotný and Prof Roman Li{\v{s}}{\v{c}}{\'a}k at Na Homolce Hospital (Prague, Czech Republic) for kindly sharing the anonymized patient data used in this study. This research was funded in part by Sweden’s Innovation Agency (Vinnova), grant number 2022-03023, and supported by the Wallenberg AI, Autonomous Systems and Software Program (WASP) funded by the Knut and Alice Wallenberg Foundation.

\section*{Author Contribution Statement}
\textbf{Joakim da Silva}: Formal analysis (supporting); Funding acquisition (equal); Investigation (lead); Methodology (equal); Project administration (equal); Resources (equal); Software (equal); Validation (lead); Visualization (lead); Original draft preparation (equal); Review \& editing (equal). \textbf{Daniel Hernández Escobar}: Formal analysis (lead); Investigation (supporting); Methodology (supporting); Software (equal), Validation (supporting); Original draft preparation (equal); Review \& editing (equal). \textbf{Tor Kjellsson Lindblom}: Data curation (lead); Investigation (supporting); Resources (equal); Software (supporting); Review \& editing (supporting). \textbf{Håkan Nordström}: Conceptualization (supporting); Data curation (supporting); Formal analysis (supporting); Funding acquisition (equal); Methodology (equal); Original draft preparation (supporting); Review \& editing (supporting). \textbf{Jens Sjölund}: Conceptualization (lead); Funding acquisition (equal); Methodology (supporting); Project administration (equal); Original draft preparation (equal); Review \& editing (equal). 

\section*{Conflict of Interest Statement}
Joakim da Silva, Tor Kjellsson Lindblom and Håkan Nordström are employees of Elekta, the company which manufactures the Leksell Gamma Knife and develops Leksell GammaPlan.

\clearpage

\section*{Appendices}
\addcontentsline{toc}{section}{\numberline{}Appendix}
\renewcommand{\thesubsection}{\Alph{subsection}}

\subsection{Explicit problem statement}\label{app:dual_lp}
Sjölund et al.\cite{sjolund2019linear} formulate a linear program for inverse planning in Gamma Knife radiosurgey and then describe how its solution can be obtained more efficiently by finding and solving the corresponding \emph{dual} linear program, given as follows
\begin{equation*}
\begin{aligned}
& \underset{x}{\text{minimize}}  
& & c^\top x \\
& \text{subject to}
& &  A x \leq b,\\
& & & \ell \leq x \leq u.
\end{aligned}
\end{equation*}
\noindent where
\begin{align}
    c^\top &= 
\begin{pmatrix}
-1 & -1 & 1 & 1 & 1 & 0
\end{pmatrix},\\
A &= 
\begin{pmatrix}
\varphi^\top_{\text{TS}} & \varphi^\top_{\text{TI}} & -\varphi^\top_{\text{R}} & -\varphi^\top_{\text{LD}} & -\varphi^\top_{\text{O}} & -C^\top_{\text{S}} \\
0 & 0 & 0 & 0 & 0 & T^\top
\end{pmatrix},\\
b^\top &= 
\begin{pmatrix}
    0 & w_{\text{BOT}}
\end{pmatrix}\\
\ell^\top &= (0 \quad 0 \quad 0 \quad 0 \quad 0 \quad 0),\\
u^\top &= 
\begin{pmatrix}
\frac{w_{\text{TS}}}{N_{\text{TS}}} & 
\frac{w_{\text{TI}}}{N_{\text{TI}}} & 
\frac{w_{\text{S}}}{N_{\text{S}}} & 
\frac{w_{\text{LD}}}{N_{\text{LD}}} & 
\infty &
\infty
\end{pmatrix}.
\end{align}

In the above equations, a superscript $\top$ is used to denote matrix transpose, and vector and matrix entries are themselves sub-vectors and matrices of relevant sizes. Subscripts TS, TI, S, G and O denote entries related to the dose points for the target surface, target interior, ring, low-dose volume, and OAR, respectively. $w_\alpha$ is the objective function weight of the primal problem for $\alpha \in \{\text{TS}, \text{TI}, \text{R}, \text{LD}, \text{BOT}\}$ and $N_\alpha$ is the number of dose points in set $\alpha \in \{\text{TS}, \text{TI}, \text{R}, \text{LD}\}$. Further,
\[
\varphi_{\text{TS}} = \frac{\phi_{\text{TS}}}{D_\text{TS}}, \quad 
\varphi_{\text{TI}} = \frac{\phi_{\text{TI}}}{D_\text{TI}}, \quad 
\varphi_{\text{R}} = \frac{\phi_{\text{R}}}{D_\text{R}}, \quad 
\varphi_{\text{LD}} = \frac{\phi_{\text{LD}}}{D_\text{LD}}, \quad 
\varphi_{\text{O}} = \frac{\phi_{\text{O}}}{D_\text{O}}, \quad 
C_{\text{S}} = \frac{C}{D_\text{T}/\varphi_{\text{cal}}},
\]
where $D_\text{T}$ is the target prescription dose, $D_\text{O}$ is the OAR max dose, $\phi_\alpha$ for $\alpha \in \{\text{TI}, \text{TS}, \text{R}, \text{LD},  \text{O}\}$ is the dose influence matrix for the dose points in the set $\alpha$, $D_\alpha$ for $\alpha \in \{\text{TS}, \text{TI}, \text{R}, \text{LD}\}$ is the threshold dose for dose points in the set $\alpha$,
\begin{align}
    C &= I_{N_\text{iso}}\otimes I_8 \otimes 1_{1\times 3}\\
    T &= I_{N_\text{iso}}\otimes 1_{8\times 1},
\end{align}
$I_n$ is the identity matrix of size $n\times n$, $1_{m\times n}$ is an $m\times n$ matrix of all ones, and $\otimes$ denotes the Kronecker product.


In the presented implementation, we let $D_\text{TS} = D_\text{R} = D_\text{T}$, and $w_\text{TS}=w_\text{TI} = w_\text{T}$. We use two separate low-dose thresholds, $D_{1,\text{LD}}$ and $D_{2,\text{LD}}$, resulting in two low-dose volumes, $[D_{1,\text{LD}}, D_{1,\text{LD}} + \Delta)$ and $[D_{2,\text{LD}}, D_{2,\text{LD}} + \Delta)$, in both the first and second passes, in accordance with the clinical software.

We extend the above problem to cases with $n>1$ targets, indexed by $k \in \{1, \dots, n\}$, by replacing each expression involving subscripts $\text{TS}$, $\text{TI}$, and $\text{R}$ with $n$ expressions with subscripts $k,\text{TS}$, $k,\text{TI}$, and $k,\text{R}$. In the definition of $C_\text{S}$, we replace $D_\text{T}$ by the mean target prescription dose $D_{T,\text{mean}}$.
Further, we let
\[
w_{k,\text{TS}} = q_{k,\text{A}} w_{\text{TS}}, \quad
w_{k,\text{TI}} = q_{k,\text{V}} w_{\text{TI}}, \quad
w_{k,\text{R}} = \frac{w_\text{R}}{n},
\]
where $q_{k,\text{A}}$ and $q_{k,\text{V}}$ are the fractions of the total target area and volume contributed by target $k$, respectively. For multiple-OAR cases, each expression involving the subscript $O$ is replaced by one such expression for each OAR.

\section*{References}
\vspace*{-15mm}






\bibliographystyle{medphy}    


\end{document}